# Data Fusion for Full-Range Response Reconstruction via Diffusion Models

**Wingho Feng[a], Quanwang Li[a], Chen Wang[a,*], Jian-sheng Fan[a]**

## Abstract

Accurately capturing the full-range response of structures is crucial in structural health monitoring (SHM) for ensuring safety and operational integrity. However, limited sensor deployment due to cost, accessibility, or scale often hinders comprehensive monitoring. This paper presents a generative data fusion framework utilizing diffusion models, to reconstruct the full-range structural response from sparse and heterogeneous sensor measurements. We incorporate Diffusion Posterior Sampling (DPS) into the reconstruction framework, using sensor measurements as probabilistic constraints to guide the sampling process. Three forward models are designed: Direct Observation Mapping (DOM), Channel-based Observation Mapping (COM), and Neural Network Forward Model (NNFM), enabling flexible adaptation to different sensor placement conditions and reconstruction targets. The proposed framework is validated on a steel plate shear wall exhibiting nonlinear responses. By simultaneously sampling 100 realizations and averaging them as the ensemble prediction result, the three forward models achieve Weighted Mean Absolute Percentage Errors of 1.62% (DOM), 3.27% (COM), and 3.49% (NNFM). Sensitivity analyses further demonstrate robust performance under varying hyperparameters, sensor configurations, and noise levels. The proposed framework shows new possibilities for probabilistic modeling and decision-making in SHM by harnessing the capabilities of diffusion models, offering a novel data fusion approach for full-range monitoring of structures.

## 1. Introduction

In structural health monitoring (SHM), accurately capturing the full-range response of a structure is paramount for ensuring both its safety and operational integrity [1],[2]. However, practical constraints, such as limited access, high sensor installation costs, or the vast scale of structures, often restrict the number of sensors that can be deployed [3]-[5]. Consequently, data obtained directly from sensors may be insufficient to provide comprehensive information, necessitating data fusion algorithms to bridge the gap between them, assisting in decision-making [6],[7].

Inferring full-range behavior from local measurements is an ill-posed problem, which a solution either does not exist or not unique [8]-[10]. For instance, in a 1/10-scale nuclear containment structure, even with hundreds of strain gauges deployed for crack monitoring over an area of 180 $m^2$ during pressure tests, these sensors provide limited coverage, making it difficult to detect damage in unmonitored regions. This incomplete data means multiple structural states could yield similar sensor readings, creating ambiguity in identifying the true state. Such ambiguity increases the risk of undetected damage, inaccurate health assessments, and potential structural failure or safety hazards. This highlights the need for advanced data fusion techniques to reconstruct full-range structural responses from sparse sensor data, enabling engineers to comprehensively evaluate structural health and identify potential failure points.

[a] Department of Civil Engineering, Tsinghua University, Beijing, 100084, China.
* Corresponding Author: Email:. chwang@tsinghua.edu.cn.

Researchers have developed various data fusion approaches for full-range response reconstruction, with model-based methods dominating in the 2010s. These methods estimate external loads or model parameters from sensor data. Mainstream approaches rely on the Kalman filter and its variations, which estimate system states through an iterative two-step process: prediction and update. The prediction step within the Kalman filter is typically deterministic. These techniques include Tikhonov regularization [11], joint-state estimation [12], principal component analysis (PCA) [13], partial least squares regression [14], and combinations of singular value decomposition (SVD) with pseudo-inverse operators [15] for state estimation with measurements. However, the Kalman filter-based methods often assume a fixed stiffness matrix and static conditions, making them unsuitable for modeling nonlinear structural behavior. Other approaches treat external loads or model parameters as random variables to incorporate probabilistic uncertainty. Narouie et al. introduced the statistical finite element method (statFEM), using Bayesian updates to infer displacements from sparse measurements, enhancing computational efficiency while addressing model-reality mismatches [16]. Li et al. proposed a sparse Bayesian estimation method to identify input loads and reconstruct complete structural responses by iteratively estimating force locations and time histories, effectively handling uncertainties in both input and measurement data [17]. Although these methods capture complex uncertainties, they can be computationally demanding, especially when large matrix operations and iterative processes are required for extensive structures.

In contrast, machine learning (ML [18], including deep learning, DL [19])-based methods represent an alternative category of data fusion methods for full-range response reconstruction. These approaches learn complex, often nonlinear and high-dimensional relationships directly from data, bypassing the need for analysis based on a predefined physical model. A straightforward ML-based approach is to directly learn the mapping from sensor data to responses at specific points of interest. Existing studies have applied ML models, including temporal Convolutional Neural Networks (CNNs) [20],[21], Long Short-Term Memory (LSTM) networks [22], and GRU [23] or their combination and modification, to reconstruct response of unmeasured positions. However, these methods generally focus on predicting responses at specific locations rather than across entire structures, and extending them to full-range response reconstruction may require substantially larger neural networks. Furthermore, these approaches often neglect the critical spatial relationships between sensors, which are essential for full-range response reconstruction.

To consider the spatial relationship, researchers have employed CNN-based methods, treating field elements as image pixels, to leverage computer vision techniques for field-related problems. This perspective has been widely applied in computational fluid dynamics [24]-[26]. For instance, Guo et al. proposed a CNN model using the Signed Distance Function to predict steady laminar flow fields, achieving 4,000× faster predictions with minimal error [27]. Similarly, Bhatnagar et al. developed a CNN model for velocity and pressure predictions in unseen flow conditions, delivering near-real-time performance across airfoil geometries [28]. More recently, CNNs have been explored for structural analysis that can consider spatial relationship. Nie et al. introduced a CNN-based method for predicting stress fields in 2D linear elastic cantilevered structures under static loads, treating geometry, loads, and boundary conditions as image channels [29]. Jadhav et al. [30] and Jiang et al. [31] extended Nie's methods within diffusion and GAN frameworks, enabling CNNs to address more complex problems. While these methods effectively address forward problems by leveraging spatial relationships, they are typically limited to deterministic solutions, restricting their capacity to capture uncertainty in inverse problems. Additionally, significant challenges remain in incorporating measurements, particularly sparse and heterogeneous data, to guide inference in full-range response reconstruction tasks.

In this study, we propose a novel data fusion approach to combine incomplete, heterogeneous measurements for full-range response reconstruction via diffusion models. We first introduce the preliminaries of diffusion models and detail our methodology based on Diffusion Posterior Sampling (DPS) for integrating sensing data. Next, we validate the proposed method on a benchmark steel plate shear wall experiment. We analyze its accuracy, robustness and inference speed with respect to different hyperparameters, sensor placement strategies, and data noise. Finally, we conclude with a discussion of the implications of our findings in SHM and the broader field of data fusion.

# 2. Methodology

## 2.1. Preliminaries of diffusion models

Diffusion models [32] are generative models designed to capture complex data distributions by progressively adding noise to data and then learning to reverse this process. The forward process corrupts the data with Gaussian noise over multiple steps, while the reverse process, parameterized by a neural network, removes the noise step by step to reconstruct the original data. This approach allows diffusion models to generate realistic data samples by modeling intricate and high-dimensional distributions, i.e. implicitly draw samples from complex response distributions.

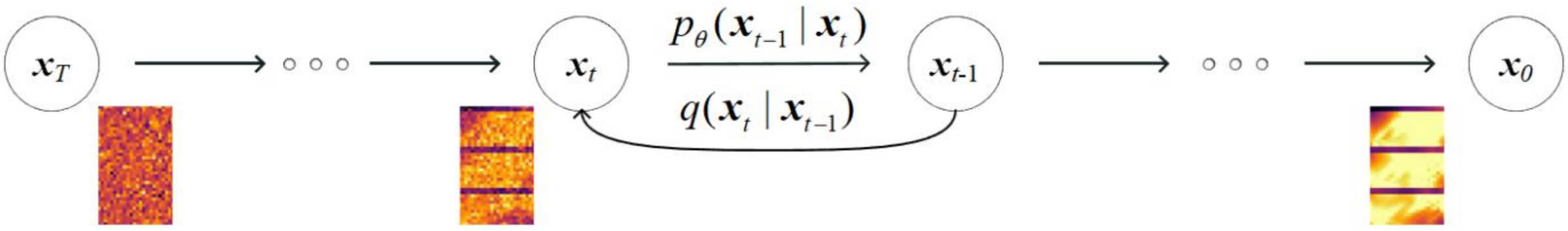

**Fig 1**.The probabilistic graphical model of Diffusion Models

Both the diffusion forward process and denoising process are Markov chain. The diffusion trajectory is denoted as $(\boldsymbol{x}_0, \ldots, \boldsymbol{x}_t, \ldots, \boldsymbol{x}_T)$, where $\boldsymbol{x}_0$ represents the original noise-free data, $\boldsymbol{x}_T$ denotes the terminal noisy variable, which approaches an isotropic standard Gaussian, and $\boldsymbol{x}_t$ indicates the intermediate variable at the $t$-th step. The transition kernel and the evolution of its parameters can be expressed as:

$$q(\mathbf{x}_t | \mathbf{x}_{t-1}) = \mathcal{N}(\sqrt{1-\beta_t}\mathbf{x}_{t-1}, \beta_t \mathbf{I}), \qquad \mathbf{x}_t = \sqrt{1-\beta_t}\mathbf{x}_{t-1} + \sqrt{\beta_t}\boldsymbol{\varepsilon}_t \tag{1}$$

where $\mathcal{N}$ denotes the Gaussian distribution, $\beta_t \in (0,1)$ is the noise variance at step $t$, which controls the extent of isotropic Gaussian perturbation, and $\boldsymbol{\varepsilon}_t$ is a Gaussian random variable sampled accordingly. Defining $\alpha_t = 1-\beta_t$ and $\bar{\alpha}_t = \prod_{i=1}^{t}\alpha_i$, it follows that

$$q(\boldsymbol{x}_t \mid \boldsymbol{x}_0) = \mathcal{N}(\boldsymbol{x}_{t-1}; \sqrt{\bar{\alpha}_t}\boldsymbol{x}_0, (1-\bar{\alpha}_t)\mathbf{I}), \qquad \boldsymbol{x}_t = \sqrt{\bar{\alpha}_t}\boldsymbol{x}_0 + \sqrt{1-\bar{\alpha}_t}\boldsymbol{\varepsilon}_t \tag{2}$$

The above expressions indicate that in the forward diffusion process, any intermediate state $\boldsymbol{x}_t$ can be directly obtained from the original data $\boldsymbol{x}_0$ by a single sampling step, thereby simplifying the training procedure for constructing samples. Moreover, as the diffusion step increases, the structural information embedded in the original data is progressively corrupted by noise, eventually converging to an isotropic Gaussian distribution. This property ensures that the forward process satisfies the initial condition required for subsequent sampling from a standard normal distribution.

The objective of the reverse stage is to learn the reverse diffusion trajectory from the Gaussian prior back to the original data $\boldsymbol{x}_0$. The reverse path is denoted as $(\boldsymbol{x}_T, \ldots, \boldsymbol{x}_t, \ldots, \boldsymbol{x}_0)$. Based on derivations by Anderson [34], the reverse stochastic differential equation (SDE) corresponding to Eq. (2) can be obtained as:

$$\boldsymbol{x}_{t-1} = \frac{1}{\sqrt{1-\beta_t}}\left[\boldsymbol{x}_t + \frac{\beta_t}{2}\nabla_x \log p_t(\boldsymbol{x}_t)\right] + \sqrt{\beta_t}\boldsymbol{\varepsilon}_t \tag{3}$$

where $p_t(\boldsymbol{x}_t)$ represents the probability density of the intermediate variable $\boldsymbol{x}_t$ at step $t$. The only unknown term is the score function $\nabla_{x_t} \log p_t(\boldsymbol{x}_t)$, which characterizes the gradient of the data distribution. Song et al. [34] proposed to approximate this score function using neural networks, trained with the following objective function

$$\boldsymbol{\theta}^* = \arg\min_{\boldsymbol{\theta}} \mathbb{E}_t\left\{\mathbb{E}_{\boldsymbol{x}(0)}\mathbb{E}_{\boldsymbol{x}(t)\mid\boldsymbol{x}(0)}\left[\| \boldsymbol{s}_{\boldsymbol{\theta}}(\boldsymbol{x}(t),t) - \nabla_{\boldsymbol{x}(t)} \log p_{0t}(\boldsymbol{x}(t) \mid \boldsymbol{x}(0))\|_2^2\right]\right\} \tag{4}$$

In practice, to obtain the optimal parameters of the neural network, Eq. (4) can be discretized, and the expectation of the objective function can be estimated using the Monte Carlo method. During each optimization step, $M$ time indices $t_m \in \{1,\ldots,T\}$ are uniformly sampled, together with Gaussian noise samples $\boldsymbol{\varepsilon}_m$ drawn from the standard normal distribution. Based on Eq. (2), the corresponding noisy data are constructed. It can be shown that the noise $\boldsymbol{\varepsilon}_m$ and the data distribution satisfy the following relation:

$$\nabla_{x_t} \log p_t(\boldsymbol{x}_t \mid \boldsymbol{x}_0) = \frac{-1}{\sqrt{1-\bar{\alpha}_t}}\boldsymbol{\varepsilon}_m \tag{5}$$

Accordingly, the training objective can be expressed as

$$\mathcal{L} = \frac{1}{M}\sum_{m=1}^{M} \| \boldsymbol{s}_{\boldsymbol{\theta}}(\boldsymbol{x}_{t_m}, t_m) + \frac{1}{\sqrt{1-\bar{\alpha}_{t_m}}} \boldsymbol{\varepsilon}_m \|_2^2 \tag{6}$$

In summary, the diffusion model learns an approximation to the data distribution by modeling the reverse denoising process, thereby enabling effective representation of high-dimensional data distributions.

## 2.2. Sensor fusion using DPS

Reconstructing full-range response from local measurement can be regarded as modeling $p(\boldsymbol{x} \mid \boldsymbol{y})$, where $\boldsymbol{x}$ is the full-range response and $\boldsymbol{y}$ is the local sensor measurement. The Eq. (3) introduce a method to model $p(\boldsymbol{x})$, which is not dependent on the measurement. To allow the incorporation of sensor measurements into the reconstruction process, we apply Diffusion Posterior Sampling (DPS) [35].

The key idea of DPS is replacing $\nabla_{\boldsymbol{x}} \log p(\boldsymbol{x})$ with $\nabla_{\boldsymbol{x}} \log p(\boldsymbol{x} \mid \boldsymbol{y})$ in Eq. (3), therefore

$$\boldsymbol{x}_{t-1} = \frac{1}{\sqrt{1-\beta_t}} \left[ \boldsymbol{x}_t + \frac{\beta_t}{2} \nabla_{\boldsymbol{x}} \log p_t(\boldsymbol{x}_t \mid \boldsymbol{y}) \right] + \sqrt{\beta_t} \boldsymbol{\varepsilon}_t \tag{7}$$

can sample data that follows $p(\boldsymbol{x} \mid \boldsymbol{y})$. To compute the conditional score, we decompose it using Bayesian theorem:

$$\nabla_{\boldsymbol{x}} \log p_t(\boldsymbol{x}_t \mid \boldsymbol{y}) = \nabla_{\boldsymbol{x}} \log p_t(\boldsymbol{y} \mid \boldsymbol{x}_t) + \nabla_{\boldsymbol{x}} \log p(\boldsymbol{x}_t) \tag{8}$$

where the second term $\nabla_{\boldsymbol{x}} \log p(\boldsymbol{x}_t)$ (prior term) is estimated using a pre-trained score function $\boldsymbol{s}_{\boldsymbol{\theta}}^*$, and the first term $\nabla_{\boldsymbol{x}} \log p_t(\boldsymbol{y} \mid \boldsymbol{x}_t)$ (likelihood term) incorporates sensor data into the reconstruction process. The prior score function is known, and the following focuses on deriving the likelihood term. The likelihood term can be approximated by (theoretical proved in [35])

$$\nabla_{\boldsymbol{x}} \log p_t(\boldsymbol{y} \mid \boldsymbol{x}_t) \approx \nabla_{\boldsymbol{x}} \log p_t(\boldsymbol{y} \mid \hat{\boldsymbol{x}}_0) \tag{9}$$

where $\hat{\boldsymbol{x}}_0$ are the predicted clean full-range response $\boldsymbol{x}_{\boldsymbol{0}}$. It can be directly calculated by

$$\hat{\boldsymbol{x}}_0(\boldsymbol{x}_t) \simeq \frac{1}{\sqrt{\bar{\alpha}_t}} (\boldsymbol{x}_t + (1-\bar{\alpha}_t) \boldsymbol{s}_{\boldsymbol{\theta}}^*(\boldsymbol{x}_t, t)) \tag{10}$$

The relationship between $\boldsymbol{x}_0$ and $\boldsymbol{y}$ are determined by a forward model. In this case, assuming the measurement contains a Gaussian noise, then

$$\boldsymbol{y} = \mathcal{A}(\boldsymbol{x}_0) + \boldsymbol{n}, \qquad \boldsymbol{n} \sim \mathcal{N}(0, \sigma^2 \mathbf{I}) \tag{11}$$

where $\mathcal{A}(\cdot): \mathbb{R}^D \mapsto \mathbb{R}^N$ is a differentiable operator that maps $\boldsymbol{x}_0 \in \mathbb{R}^D$ to $\boldsymbol{y} \in \mathbb{R}^N$. We will discuss the design of the forward model $\mathcal{A}(\cdot)$ in the following section. The random variable $\boldsymbol{n}$ denotes that the noise of measurement follows Gaussian distribution. Other description of noise, such as Poisson noise, can be analyzed similarly to quantify the uncertainty of measurement. The likelihood is given by

$$p_t(\boldsymbol{y} \mid \hat{\boldsymbol{x}}_0) = \frac{1}{\sqrt{(2\pi)^N \sigma^{2N}}} \exp\left[-\frac{\| \boldsymbol{y} - \mathcal{A}(\hat{\boldsymbol{x}}_0)\|_2^2}{2\sigma^2}\right] \tag{12}$$

The likelihood term, representing the contribution of sensor data, is:

$$\nabla_{\boldsymbol{x}} \log p_t(\boldsymbol{y} \mid \hat{\boldsymbol{x}}_0) = -\frac{1}{\sigma^2} \nabla_{\boldsymbol{x}_t} \| \boldsymbol{y} - \mathcal{A}(\hat{\boldsymbol{x}}_0)\|_2^2 \tag{13}$$

which is tractable once the forward model $\mathcal{A}(\cdot)$ is determined. The conditional score function therefore can be written as

$$\nabla_{\boldsymbol{x}} \log p_t(\boldsymbol{x}_t \mid \boldsymbol{y}) \approx \boldsymbol{s}_\theta^*(\boldsymbol{x}_t, t) - \rho \nabla_{\boldsymbol{x}_t} \| \boldsymbol{y} - \mathcal{A}(\hat{\boldsymbol{x}}_0)\|_2^2 \tag{14}$$

where $\rho = \frac{1}{\sigma^2}$ is hyperparameter that embody the uncertainty of the measurement. After the conditional score function is obtained by (14), samples can be generated under measurement conditions using Eq. (7). The geometric interpretation is given in **Fig 2**. The decoupling of the forward model and score estimator allows sensor quantity and placement changes to impact only the forward model without retraining the score estimator, facilitating engineering implementations.

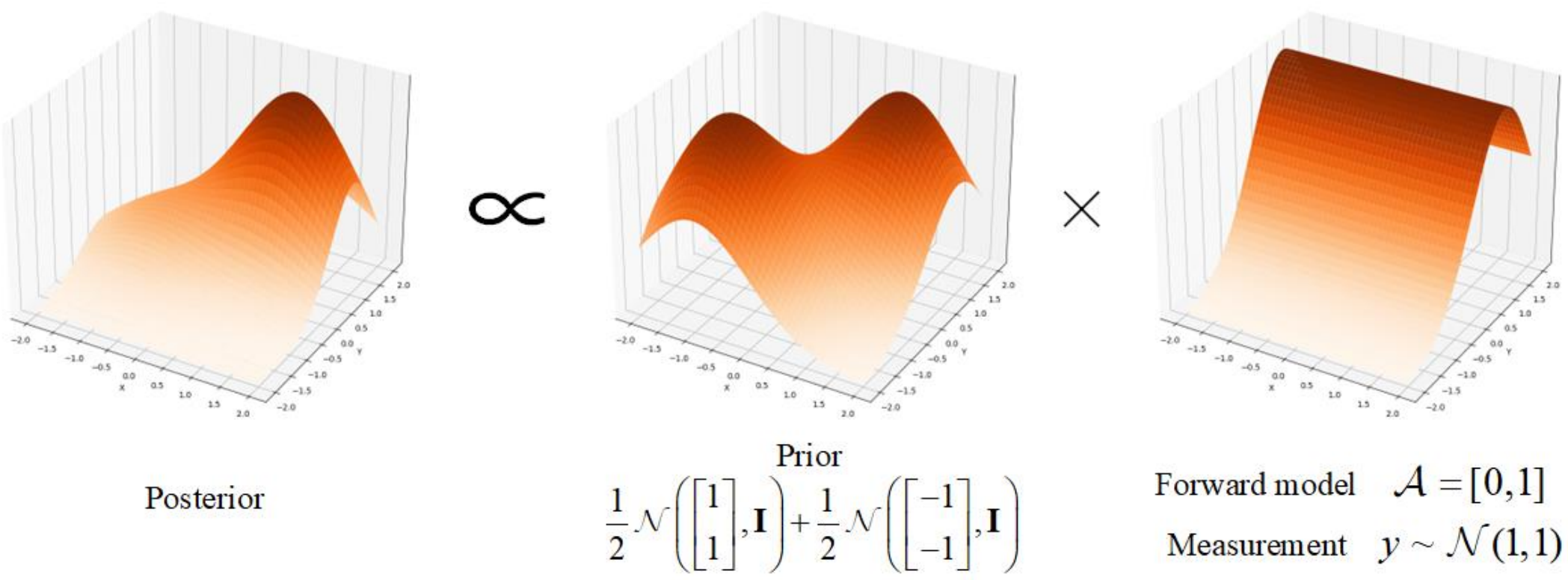

**Fig 2.** A geometric representation of DPS. Two random variables, $x$ and $y$, are used: the prior distribution is bimodal with means at [1, 1] and [-1, -1], while the likelihood is based on measurements with $y \sim \mathcal{N}(1,1)$.

For **heterogeneous** sensor measurements that collect different mechanical field data, let $\boldsymbol{Y} = \{\boldsymbol{y}_1, \boldsymbol{y}_2, \ldots, \boldsymbol{y}_K\}$ represent measurements from different sensor types. Each $\boldsymbol{y}_i$ can be modeled as:

$$\boldsymbol{y}_i = \mathcal{A}_i(\boldsymbol{x}_0) + \boldsymbol{n}_i, \qquad \boldsymbol{n}_i \sim \mathcal{N}(0, \sigma_i^2 \mathbf{I}), \tag{15}$$

where $\mathcal{A}_i$ represents the mapping operator for the *i*-th sensor type. The conditional score for heterogeneous sensors becomes:

$$\nabla_{\boldsymbol{x}} \log p_t(\boldsymbol{x}_t \mid \boldsymbol{Y}) \approx s_\theta^*(\boldsymbol{x}_t, t) - \sum_{i=1}^{K} \rho_i \nabla_{\boldsymbol{x}_t} \| \boldsymbol{y}_i - [\mathcal{A}_i(\boldsymbol{x}_0)]\|_2^2, \tag{16}$$

This extension of DPS enables robust fusion of heterogeneous measurements, facilitating accurate and comprehensive full-range response reconstruction even in complex, noisy scenarios. The complete algorithm is given in **Fig 3**. In the algorithm, $\zeta_i$ can be selected based on $\rho_i$, i.e. a larger $\zeta_i$ corresponds to sensors with higher noise levels.

**Algorithm 1** Diffusion Posterior Sampling

**Require:**
$T$, ▷ Number of steps
$s_\theta^*, \{\beta_t\}_{t=1}^T, \{\bar{\alpha}_t\}_{t=1}^T$, ▷ Prior Term Hyperparameters
$\{\zeta_i\}_{t=1}^K, \{y_i\}_{i=1}^K, \{\mathcal{A}_i\}_{i=1}^K$, ▷ Likelihood Term Hyperparameters
1: $x_T \leftarrow \mathcal{N}(\mathbf{0}, \mathbf{I})$
2: **for** $t = T$ **to** 1 **do**
3: $s \leftarrow s_\theta^*(x_t, t)$
4: $z_t \leftarrow \mathcal{N}(\mathbf{0}, \mathbf{I})$
5: $x_{t-1} \leftarrow \frac{1}{\sqrt{1-\beta_t}}[x_t + \frac{\beta_t}{2} s] + \sqrt{\beta_t} z_t$
6: $\hat{x}_0 \leftarrow \frac{1}{\sqrt{\bar{\alpha}_t}}(x_t + (1-\bar{\alpha}_t)s)$
7: $x_{t-1} \leftarrow x_{t-1} - \sum_{i=1}^K \zeta_i \nabla_{x_t} \|y_i - \mathcal{A}_i(\hat{x}_0)\|^2$
8: **end for**

**Fig 3.** DPS for measurement with Gaussian noise

## 2.3. Forward Model Design under the DPS Framework

Within the DPS framework, the forward model characterizes the mapping between structural responses and sensor observations (see Eq. (11)), which also provides the basis for computing the likelihood term. By constructing this mapping, local sensor measurements can be incorporated into the reconstruction process, enabling the diffusion model to sample from the posterior distribution. Compared to solving inverse problems, forward problems are more tractable due to their causal structure. This constitutes a key advantage of the DPS framework: by transforming ill-posed inverse problems into well-posed forward formulations, the overall modeling stability is significantly improved.

Depending on application requirements, various forms of forward models can be constructed. In this study, we introduce three design strategies for different scenarios as below.

### 2.3.1. Direct Observation Mapping (DOM)

The Direct Observation Mapping (DOM) represents the simplest form of forward modeling. It is applicable when the sensor observation variables are of the same type as the target responses to be reconstructed (e.g., stress measurements used for stress reconstruction). In this approach, a selection matrix is employed to extract the values at corresponding sensor locations. Let $\boldsymbol{x} \in \mathbb{R}^n$ denote the global structural response, and $\boldsymbol{y} \in \mathbb{R}^m$ ($m < n$) denote the sensor observations. DOM can be expressed as

$$\boldsymbol{y} = S\boldsymbol{x} \tag{17}$$

where $S \in \{0,1\}^{m \times n}$ is a fixed selection matrix that extracts the components of $\boldsymbol{x}$ corresponding to the sensor locations. As no network training is required, DOM is computationally efficient and numerically stable, making it a suitable baseline for assessing the performance of more sophisticated methods. However, its applicability is restricted to cases where the observed and reconstructed variables are identical, and it cannot be directly extended to scenarios involving heterogeneous variable types.

### 2.3.2. Channel-based Observation Mapping (COM)

The Channel-based Observation Mapping (COM) extends the applicability of forward models to cases where the observed and target variables are not of the same type. In this formulation, the observable and target responses are treated as distinct channels of a unified representation. For instance, in structural monitoring, strain measurements may be available while the objective is to reconstruct the stress field. A joint score estimator is trained to predict multiple response variables simultaneously, thereby establishing a comprehensive multi-channel representation. During sampling, the observation matrix is applied exclusively to the observable channels, which guides the reconstruction process while obviating the need for an explicit forward mapping of the target variable. COM is particularly suitable for multi-physics monitoring tasks and exhibits adaptability in scenarios involving heterogeneous observation and response variables.

### 2.3.3. Neural Network Forward Model (NNFM)

The aforementioned methods rely on fixed selection matrices or channel assumptions, lacking flexibility to capture complex mappings of structural responses. In scenarios with nonlinear coupling or heterogeneous multi-source data, selection-based methods are inadequate. Alternatively, constructing forward models purely from physical formulations faces two major limitations: (i) high computational cost, since the forward model must be repeatedly invoked during each diffusion sampling iteration, making it impractical for efficient inference; and (ii) lack of differentiability, which prevents computation of likelihood score terms. These challenges limit the direct integration of physics-based forward models into the DPS framework.

To overcome these limitations, we introduce neural networks as surrogate forward models. Neural networks are inherently differentiable, allowing efficient training via backpropagation, and provide substantially faster inference compared to physics-based models. They possess strong capability to model complex nonlinear relationships between inputs and outputs, and are particularly effective in handling heterogeneous multi-source data such as displacement, strain, and temperature measurements. More importantly, as data-driven tools, neural networks do not rely on specific physical constitutive assumptions, enabling broad applicability across diverse mapping tasks between sensor measurements and structural response fields. In practice, the neural network forward model is trained by minimizing a prediction error loss function:

$$\varphi^* = \underset{\varphi}{\operatorname{argmin}}\, \mathcal{L}(\boldsymbol{Y}, \mathcal{A}_\varphi(\boldsymbol{x})) \tag{18}$$

where $\mathcal{L}$ denotes an appropriate loss function that estimates the reconstruction error. Once trained, it can be seamlessly embedded into the diffusion posterior sampling process to achieve conditional sampling and structural response reconstruction. The overall framework of the DPS method with NNFM is illustrated in **Fig 4**.

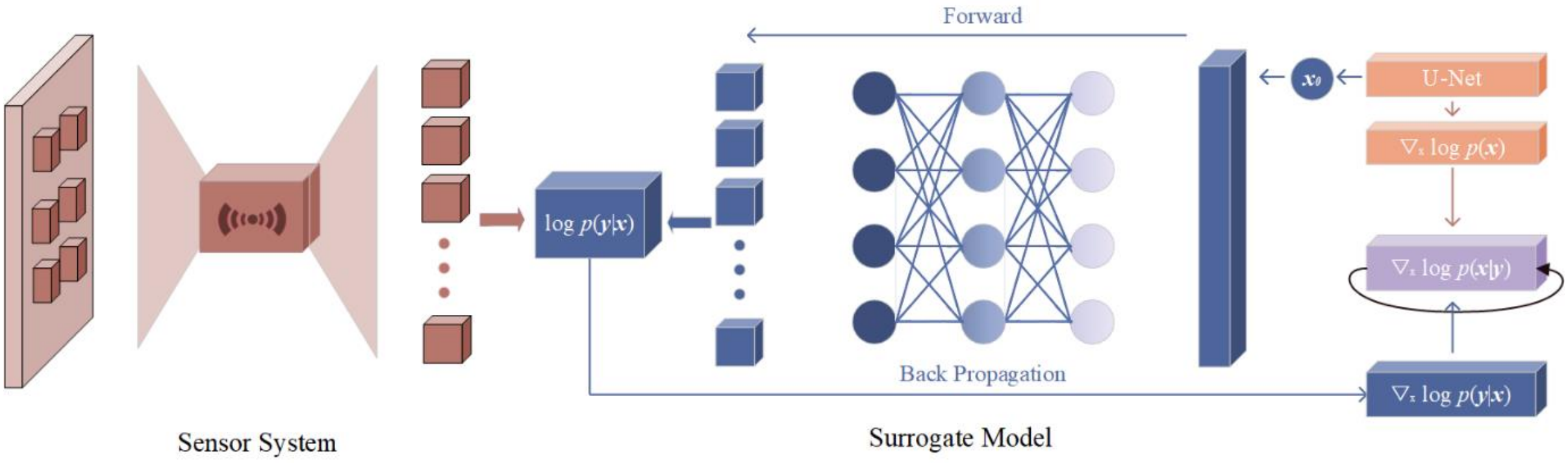

**Fig 4.** Sensor-guided sampling method via surrogate forward model

# 3. Validation

## 3.1. Case data

A steel plate shear wall (SPSW) is selected for validation, which was one of the benchmark SPSW experiments reported in the work of In-Rak Choi et al. [38], named FSPW2. A finite element model is built to simulates the structural response under displacement loading from 0 to 100 mm applied at the top left corner. The displacement range was chosen to ensure that the structure exhibits nonlinear behavior, enriching the dataset and increasing its complexity to effectively test our model. Through finite element analysis, we obtained element-wise response (strain and stress) histories throughout the loading process. We then utilized each temporal frame of the response history as an independent data point. The primary objective of this experiment is to reconstruct the full-range von Mises stress response of the specimen, given sensor measurements (of either strain or stress) at specific locations. The loading configuration and sensor placement strategy are illustrated in **Fig 5**.

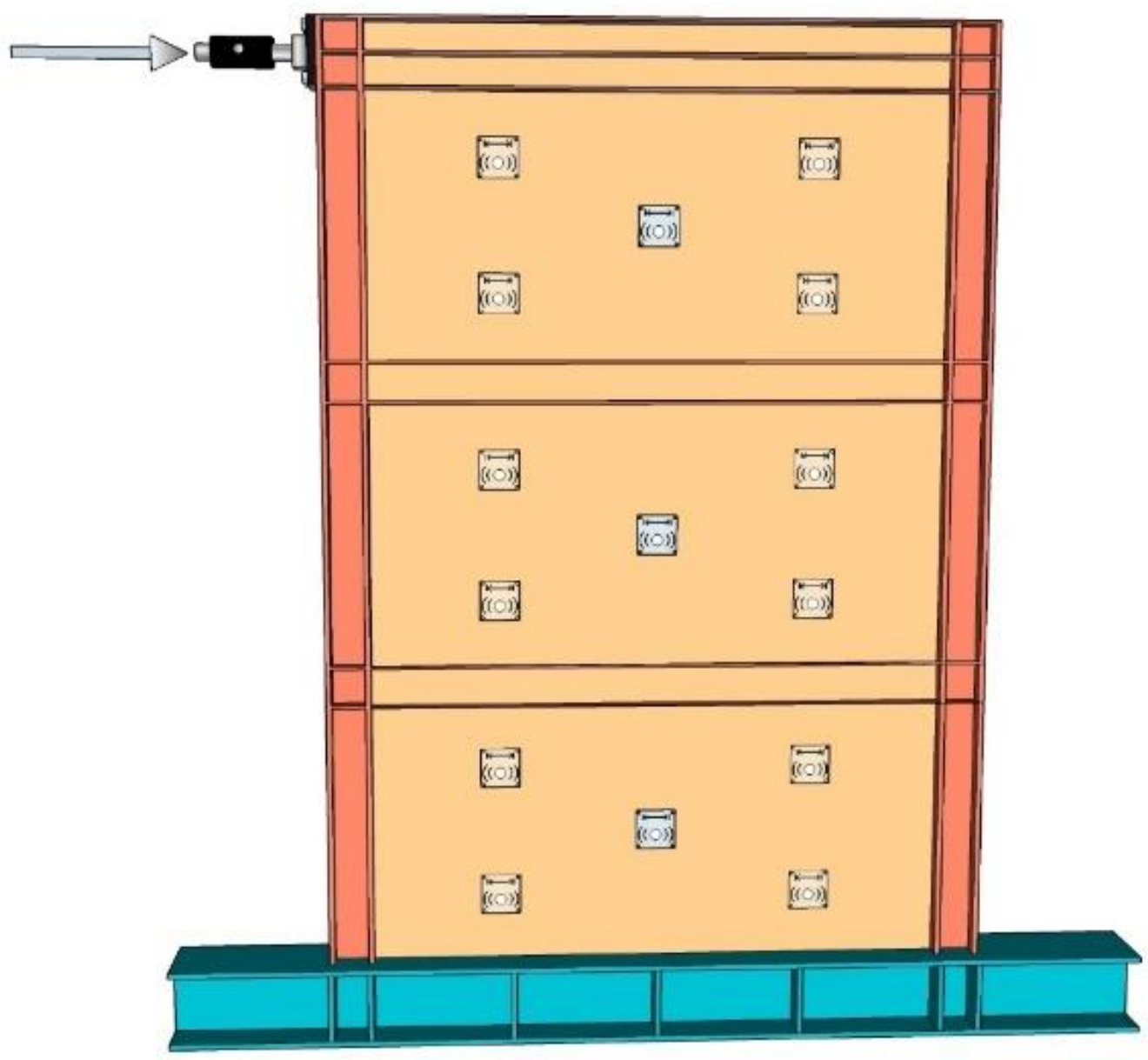

**Fig 5.** The loading sketch map of FSPW2 and its sensors placement strategy

## 3.2. Model implementation and training configuration

In diffusion models, the U-Net architecture [37] is widely employed for constructing score estimators. It is a classical CNN architecture originally developed for image denoising and segmentation tasks, with strong capabilities in feature extraction and multi-scale representation. Its architecture consists of three major components: a down-sampling module, an up-sampling module, and skip connections. This design enables the network to retain both low-level local detail features and high-level abstract semantic features, thereby enhancing its overall modeling capacity. The detailed network structure and channel dimensions are shown in **Fig 6**.

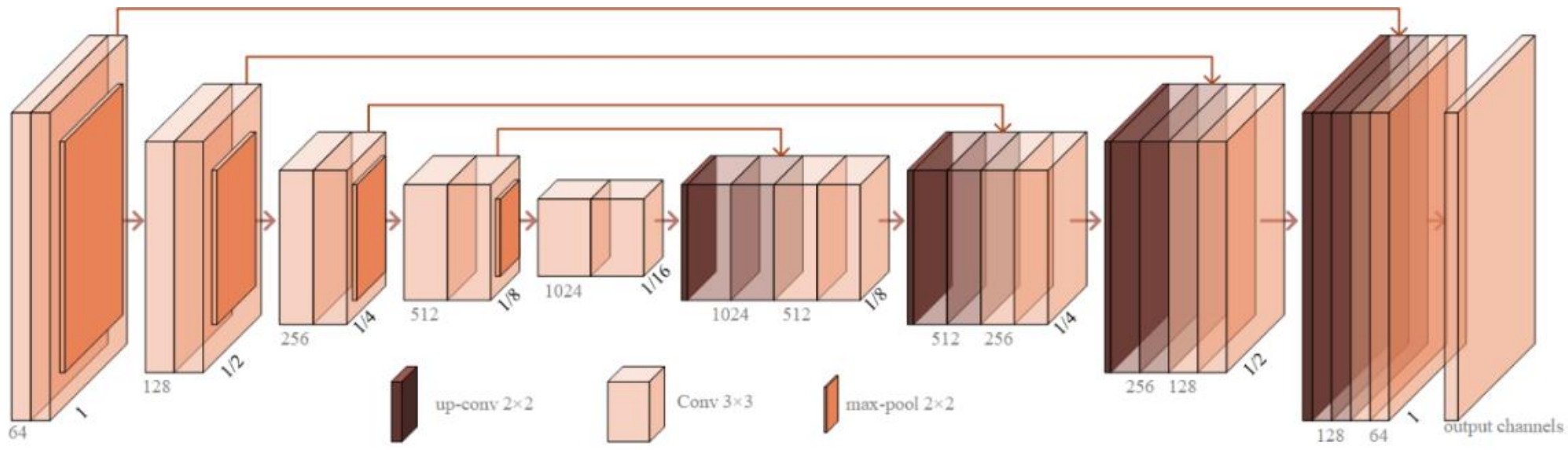

**Fig 6.** U-Net architecture and parameter configuration

Two U-Net models are trained as score estimator in this study, corresponding to the following data distributions:

(1) Single-variable model: used to model the stress field (von Mises stress), with one input channel.

(2) Multi-variable model: used to model the joint distribution of the stress and strain fields, with two input channels.

Both score estimators are optimized using the Adam optimizer with a learning rate of 0.001. Under the diffusion posterior sampling framework, different forward models are associated with different score estimators:

(1) The DOM and NNFM share the single-variable model.

(2) The COM employs the multi-variable joint model.

Furthermore, the NNFM incorporates a multilayer perceptron (MLP) to learn the mapping between the stress and strain fields. The MLP consists of an input layer (dimension 38×24), one hidden layer with 100 neurons, and an output layer (dimension 15), fully connected throughout. During training, Gaussian noise with a standard deviation of 0.1 is added to the model inputs to enhance robustness against perturbations.

Regarding the training process, we apply z-score normalization to standardize the data, which can be expressed as:

$$\boldsymbol{x}_{\text{norm}} = \frac{\boldsymbol{x} - \mu}{\sigma} \tag{19}$$

where $\mu$ and $\sigma$ represent the statistical mean and standard deviation of the data $\boldsymbol{x}$.

## 3.3. Metric for reconstruction accuracy

Based on the diffusion model–based structural response reconstruction, the task can be regarded as posterior sampling conditioned on given observations. When observational information is insufficient to fully characterize the global response of the structure, the sampled responses exhibit randomness, reflecting the inherent uncertainty in reconstruction. Therefore, from the perspective of ensemble prediction, multiple samples can be generated through the diffusion model (100 samples in this study), and their averaged value is used as the final prediction result for comprehensive evaluation. To comprehensively assess the reconstruction quality of diffusion model–generated samples, four evaluation metrics are employed.

(1) Mean Absolute Error (MAE).

$$\mathrm{MAE} = \frac{1}{N}\sum_{i=1}^{N}|\hat{x}_i - x_i| \tag{20}$$

MAE represents the average of the absolute deviations between predictions and ground truth, directly reflecting the overall bias.

(2) Root Mean Square Error (RMSE).

$$\mathrm{RMSE} = \sqrt{\frac{1}{N}\sum_{i=1}^{N}(\hat{x}_i - x_i)^2} \tag{21}$$

Compared with MAE, RMSE is more sensitive to outliers and is more suitable for penalizing large deviations.

(3) Energy Score (ES).

$$\mathrm{ES}(\hat{\boldsymbol{x}}, \boldsymbol{x}) = \mathbb{E}_{\hat{x}}\|\hat{\boldsymbol{x}} - \boldsymbol{x}\| - \frac{1}{2}\mathbb{E}_{\hat{x},\hat{x}'}\|\hat{\boldsymbol{x}} - \hat{\boldsymbol{x}}'\| \tag{22}$$

Here, $\hat{\boldsymbol{x}}$ and $\hat{\boldsymbol{x}}'$ denote two independent samples drawn from the model predictions, while $\boldsymbol{x}$ represents the ground truth. The energy score provides a comprehensive evaluation of both predictive accuracy and sample diversity, and is commonly used in ensemble forecasting tasks. Essentially, it generalizes the continuous ranked probability score (CRPS) to multivariate spaces. In this study, the absolute error is adopted as the evaluation criterion.

(4) Weighted Mean Absolute Percentage Error (WMAPE).

$$\mathrm{WMAPE} = \frac{\sum_{i=1}^{n}|x_i - \hat{x}_i|}{\sum_{i=1}^{n}|x_i|} \tag{23}$$

WMAPE uses the absolute values of the ground truth as weights. By normalizing the absolute error by the sum of the ground truth values, WMAPE eliminates the influence of magnitude and provides an intuitive interpretation of relative error.

## 3.4. Reconstruction Accuracy and Visualization

After the training of the U-Net model and the MLP used in the NNFM is completed, their performance can be validated and evaluated. All models employ the DDIM sampling strategy, with sparse sampling executed at intervals of 10 steps. The optimal hyperparameter settings differ across forward models, corresponding to their best reconstruction accuracy:

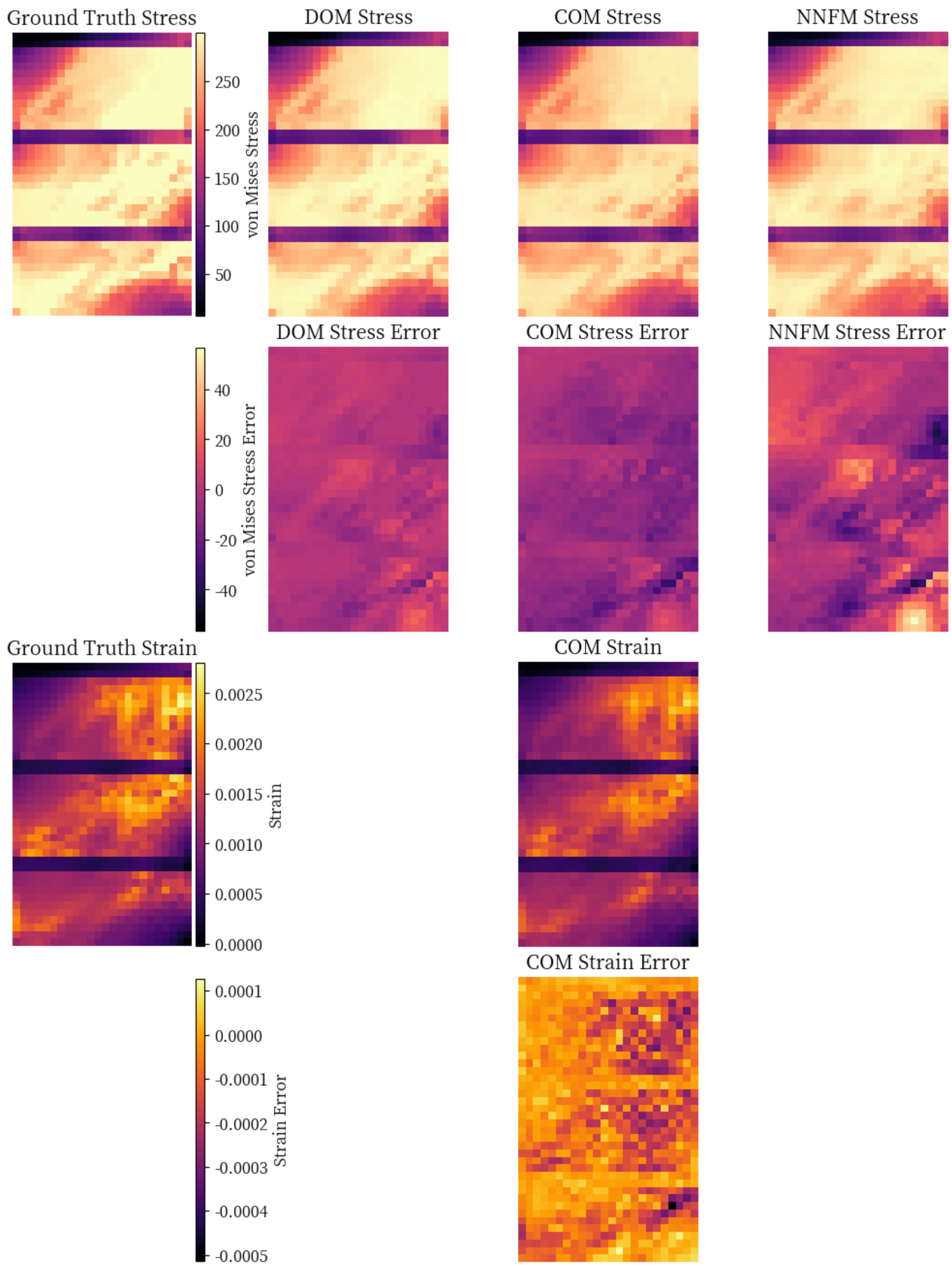

**Fig 7.** Visualization of response reconstruction and associated errors under different forward models (stress in MPa)

(1) DOM achieves its best performance with 100 sampling steps and a conditional sampling strength of 15, yielding a WMAPE of 1.62%.

(2) COM achieves its best performance with 700 sampling steps and a conditional sampling strength of 20, yielding a WMAPE of 3.27%.

(3) NNFM achieves its best performance with 700 sampling steps and a conditional sampling strength of 9, yielding a WMAPE of 3.49%.

The reconstructed stress and strain fields, along with their error distributions under optimal parameter settings, are visualized in **Fig 7**. Note that DOM and NNFM reconstruct only the stress field, without reconstructing the strain field. To ensure a fair comparison under identical observation configurations, all models utilize the same 15 sensor locations (as shown in **Fig 5**), and are trained and sampled under identical data and configuration settings. The results indicate that all three methods are capable of reconstructing the full-range response distribution of the steel plate shear wall, with generally consistent numerical agreement.

Analysis of the error distributions reveals that the three forward models exhibit relatively high errors in transition regions between high-stress and low-stress zones (e.g., in the lower-right corner), where stress gradients are pronounced. A large stress gradient corresponds to sharp spatial variations of the response, and when sensor coverage is insufficient, the models often struggle to accurately localize these variations. Even if the predicted magnitude is accurate, the spatial position may be shifted, resulting in blurring or drifting effects in the reconstruction errors. On the other hand, while COM is capable of reconstructing both stress and strain distributions, no strong correlation is observed between their respective error patterns. Both fields tend to exhibit elevated errors in regions with large gradients.

Considering the differences in reconstruction accuracy across the three forward models, DOM achieves relatively higher accuracy since its task (reconstructing global stress fields from stress observations) is comparatively simpler. In contrast, COM and NNFM reconstruct responses from strain observations. Among these, COM produces a flatter error distribution, whereas NNFM yields a more concentrated error distribution. This may be attributed to the fact that the forward MLP model in NNFM is trained solely on global fitting without explicitly encoding the spatial structure of the strain field, which leads to insufficient local modeling capability in certain regions.

## 4. Discussion

The preceding validation focused on reconstruction performance under specific parameter configurations. In this section, we extend the analysis to investigate the influence of hyperparameter settings, sensor placement strategies, and measurement noise on reconstruction accuracy, thereby offering a more comprehensive evaluation of the proposed method. In addition, the inference efficiency is assessed under varying DDIM step sizes to examine the practical applicability of the approach.

Before proceeding with the discussion, it is necessary to examine the correlation among the four metrics introduced earlier. A Pearson correlation analysis was conducted on the metric values obtained from all reconstruction results. The outcomes are presented in **Fig 8**. It can be observed that the pairwise Pearson correlation coefficients among the four metrics all exceed 0.94, indicating a very strong linear dependence. This result suggests that the metrics provide highly consistent rankings and model performance assessments, thereby allowing for a streamlined evaluation system and avoidance of redundant analyses in subsequent studies.

In summary, since the ES offers a comprehensive evaluation of ensemble predictions under uncertainty, it will be adopted as the primary metric in the following comparative analyses of different hyperparameter settings and model configurations.

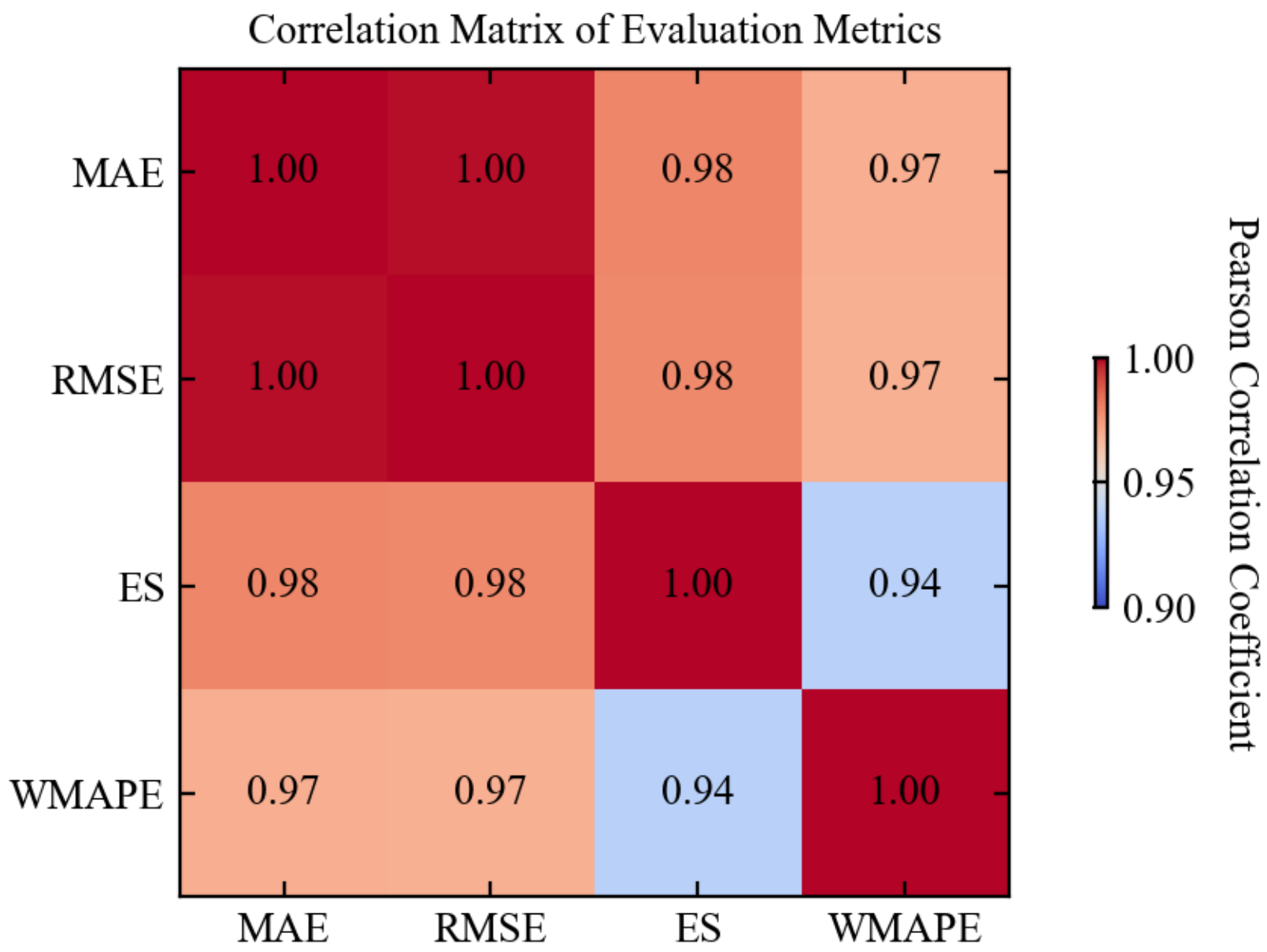

**Fig 8.** Correlation analysis of evaluation metrics (Pearson correlation coefficients)

### 4.1. Impact of hyperparameters

The sampling step number ($T$) and the conditional sampling strength ($\zeta$) are two critical hyperparameters affecting reconstruction accuracy. Specifically, the sampling step number corresponds to the number of time steps in the generative process, which can be interpreted as the discretization steps of the SDE solved by Euler–Maruyama schemes. The conditional sampling strength controls the contribution of observational information in the reverse generative process, reflecting the model's confidence weight in the measurements. **Fig 9** illustrates the effects of varying sampling steps and conditional sampling strengths on reconstruction accuracy.

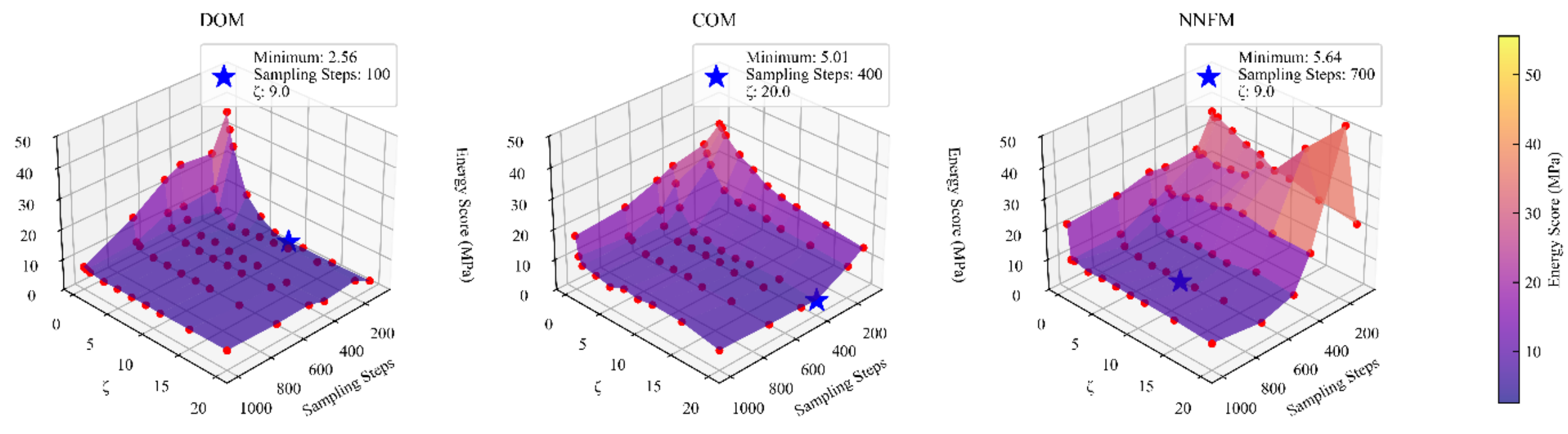

**Fig 9.** Effects of different sampling steps and conditional sampling strengths on reconstruction accuracy

Regarding the sampling step number, all three models demonstrate a general trend of improved accuracy with increasing step size. This is because longer integration horizons provide more accurate numerical approximations of the underlying SDE. In particular, DOM exhibits relatively stable reconstruction performance even under small sampling step sizes, suggesting that it can complete the response reconstruction task reliably with fewer steps. COM exhibits a smoother improvement in accuracy, whereas NNFM shows substantial fluctuations in reconstruction error at small step sizes and converges only when the step size is sufficiently large (e.g., 700). This sensitivity may stem from biases in gradient estimation by the neural network, which become amplified during sampling.

With respect to conditional sampling strength, $\zeta$ effectively controls the weight of the likelihood term $\nabla_x \log p_t(\boldsymbol{y} \mid \hat{\boldsymbol{x}}_0)$. On the one hand, insufficient $\zeta$ results in weak integration of observational information during reverse generation. On the other hand, excessive $\zeta$ may impair the numerical stability of the stochastic differential equation solver. As illustrated, when $\zeta$ approaches zero, the reconstructed accuracy decreases due to the lack of observational guidance. As $\zeta$ increases, reconstruction accuracy improves initially but tends to decline once the stability of the reverse process is compromised. Specifically, DOM and COM achieve stable accuracy across a relatively broad range of $\zeta$, whereas NNFM demonstrates higher sensitivity. For small $T$ (e.g., $T = 200$), a too large $\zeta$ increases error and destabilizes the process, as the NNFM's data-fitted gradient is biased, overweighting it amplifies and accumulates the error, making reconstructions unreliable.

The above analysis highlights the substantial impact of hyperparameter settings, particularly regarding stability estimation under small step sizes. Therefore, a systematic hyperparameter search is necessary. Complex models such as NNFM exhibit higher sensitivity, requiring more careful tuning. Through extensive hyperparameter exploration, it can be observed that optimal accuracy is attainable under proper configurations.

## 4.2. Impact of sensor placement strategies

Owing to the characteristics of the DPS framework, modifications in sensor placement can be readily incorporated by adjusting the forward model, which allows reconstruction errors to be rapidly evaluated under different configurations. This provides an efficient means of assessing the quality of alternative sensor placement schemes. Therefore, we examined reconstruction errors across various strategies to illustrate this property and to investigate the effects of different sensor configurations.

### 4.2.1. Sensor quantity

With the number of sensors treated as an independent variable, we analyzed its effect on reconstruction performance. Using the previously introduced 15-sensor configuration as the upper bound, alternative configurations with fewer sensors ( $n \in [3,15]$ ) were generated by randomly removing $15-n$ sensors. For each sensor quantity, 100 independent trials were conducted, and the average error metrics were reported to mitigate the evaluation bias caused by location variability.

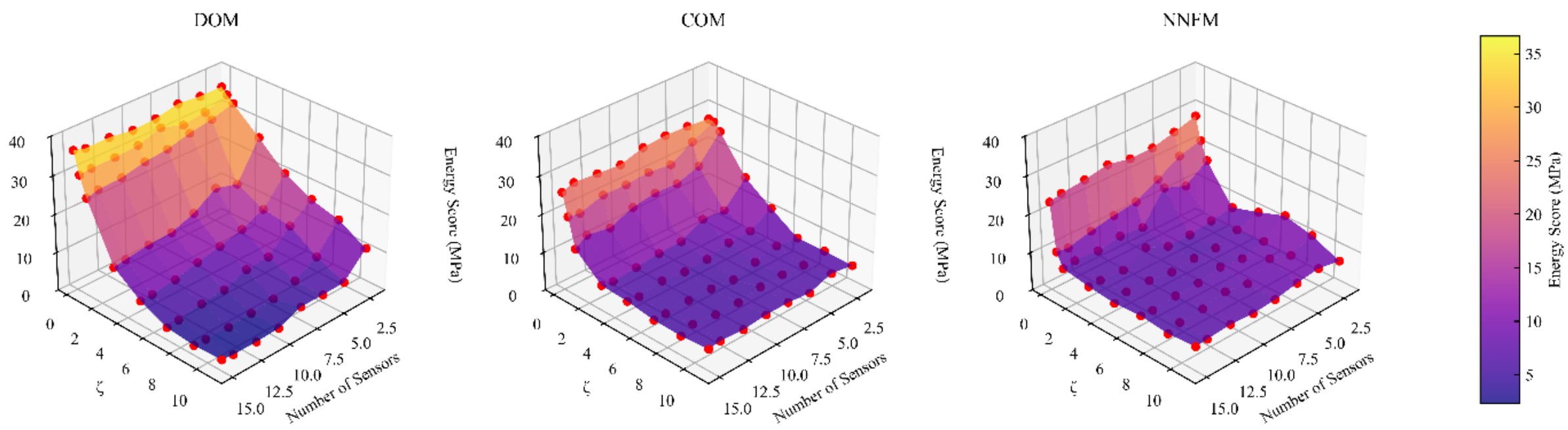

**Fig 10.** Influence of sensor quantity and conditional sampling strength on reconstruction accuracy

**Fig 10** illustrates the trends in reconstruction accuracy under varying sensor quantities and conditional sampling strengths. It can be observed that, as the number of sensors increases, the reconstruction errors of all three models decrease and eventually converge when the number of sensors becomes sufficiently large. This indicates that beyond a certain threshold, adequate observational information is already available to support accurate reconstruction. The convergence point of the error curves can be regarded as the saturation point of benefit in sensor deployment, which provides a practical reference for determining the minimum effective number of sensors in real-world applications.

### 4.2.2. Sensor positioning

To analyze the impact of sensor positioning, we employed 15 sensors across four distinct positioning strategies, as illustrated in **Fig 11**. These strategies are based on four rules: top 15 high-variance points, top 15 low-variance points, a standard placement commonly used in practice, and the random placement.

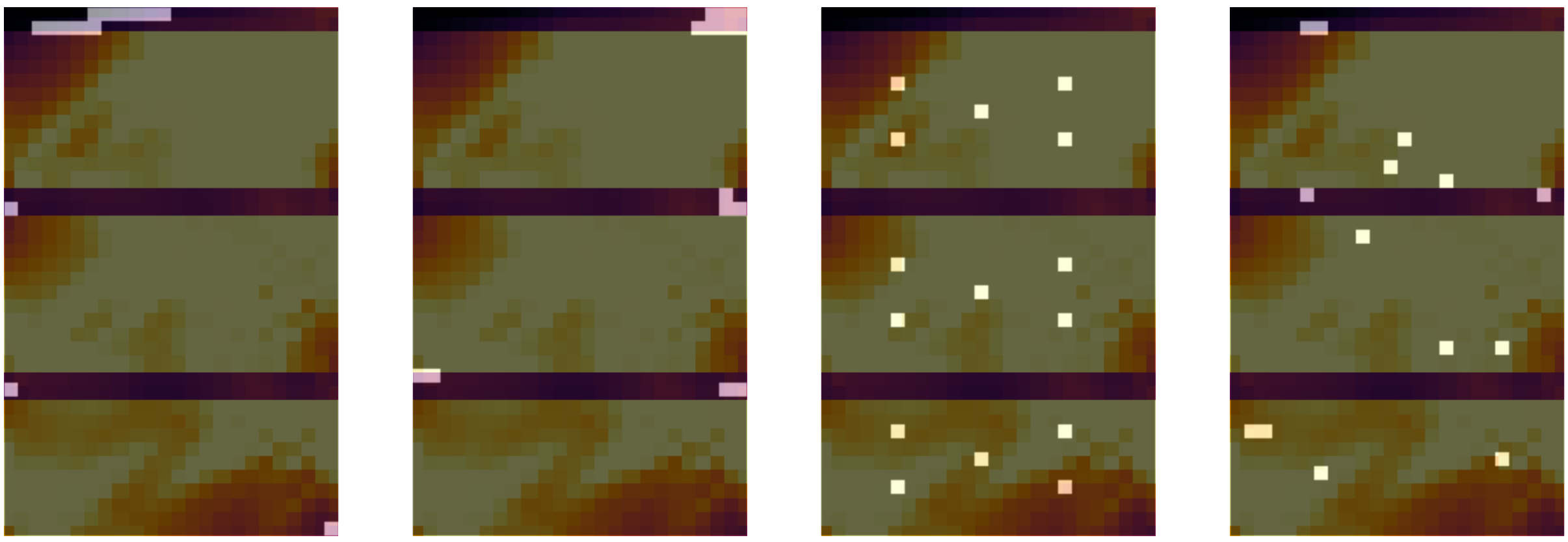

**Fig 11.** Sensor positioning (from left to right: top 15 low-variance points, top 15 high-variance points, standard placement, random placement)

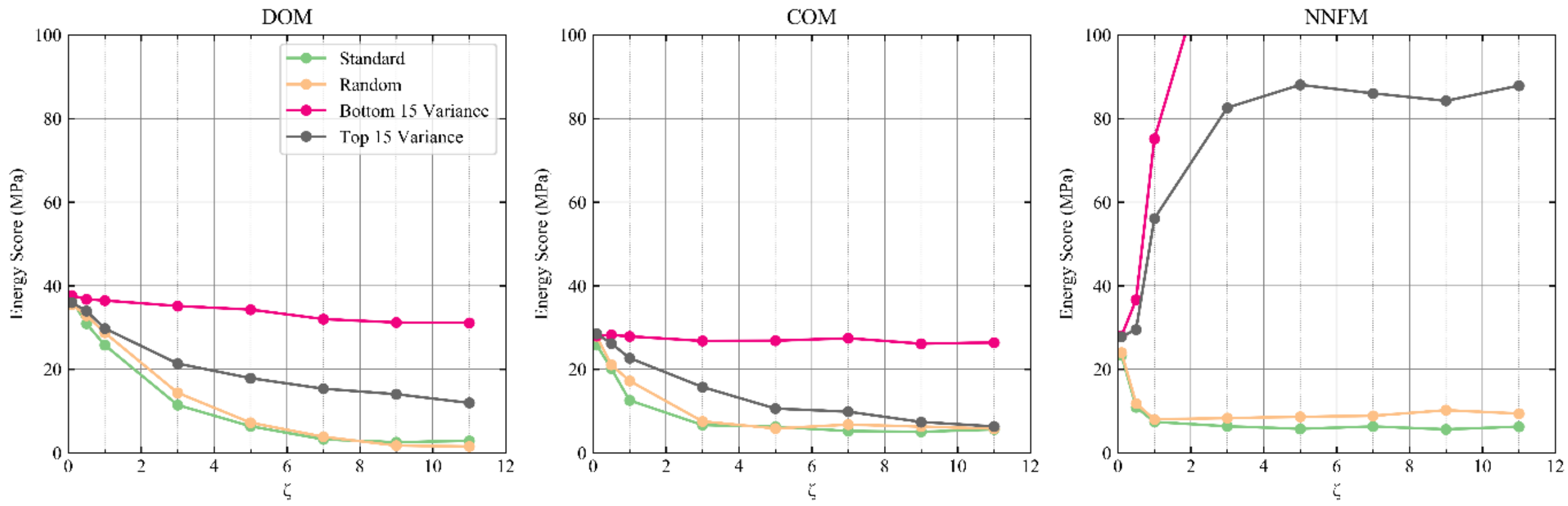

**Fig 12.** Reconstruction accuracy under different sensor placement strategies and conditional sampling strengths

The reconstruction accuracy results are presented in **Fig 12**. It is evident that all models perform worst under the minimum-variance placement, indicating that low-variance nodes provide insufficient structural response information, thereby limiting reconstruction accuracy. The maximum-variance placement, although informative, suffers from spatial concentration of sensors, which leads to redundancy and prevents optimal performance. In contrast, the standard and random placement strategies provide broader spatial coverage of the structure, yielding more representative observational information. As a result, their reconstruction accuracies are comparable and significantly superior to those of the variance-based heuristic placements.

In addition, for the NNFM, increasing the conditional sampling strength introduces pronounced fluctuations in reconstruction error, reflecting an unstable trend. This instability is associated with gradient noise and structural variability inherent in its neural network–based forward model, and suggests that under stronger observational weighting, NNFM is more sensitive to noise amplification effects.

This analysis highlights the potential of the DPS framework for optimizing sensor placement. in the present case study, we found that sensor distributions that are more spatially dispersed across the structure yield better reconstruction accuracy. Moreover, the variance of nodal responses is influential in determining performance, as it reflects the amount of information carried by each sensor.

## 4.3. Impact of data noise

Sensors in engineering applications are inevitably affected by noise, and robust algorithms are essential to mitigate this issue. Our method explicitly accounts for noise during the derivation process, making it inherently well-suited for use with noisy sensor data. In this section, we evaluate how accuracy change with different noise levels.

Specifically, noise is added to the normalized sensor measurements, with varying noise levels applied to each sample. For each noise level, we generate multiple samples with different noise that draw from a Gaussian distribution $\mathcal{N}(0,\sigma_{\text{noise}}^2\mathbf{I})$. The normalized sensor measurements have a variance of $\sigma_{\text{signal}}^2=1$. The Signal-to-Noise Ratio (SNR) is defined as the ratio of the signal variance to the noise variance, as follows:

$$\text{SNR}=10\log_{10}\left(\frac{\sigma_{\text{signal}}^2}{\sigma_{\text{noise}}^2}\right) \tag{24}$$

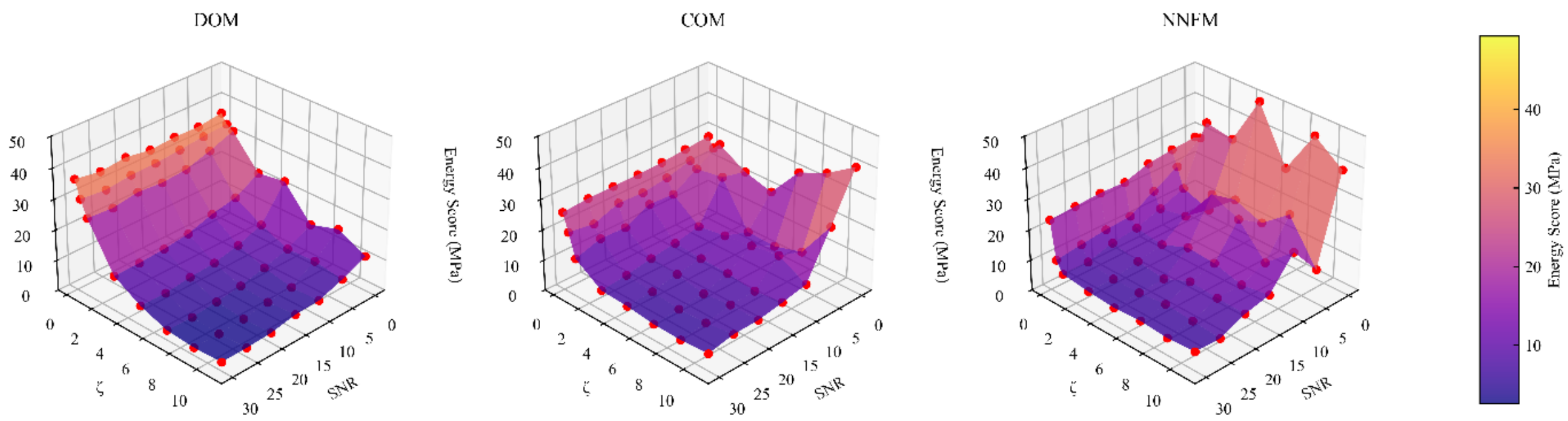

**Fig 13.** Reconstruction accuracy under different SNR and conditional sampling strengths

The results (**Fig 13**) illustrates the reconstruction accuracy trends of different forward models under varying SNR and conditional sampling strengths. The results reveal a consistent pattern across all three models (DOM, COM, and NNFM): reconstruction errors decrease markedly as the SNR increases, indicating that higher-quality observational data yield more reliable diffusion sampling. Furthermore, in the COM and NNFM models, an additional phenomenon is observed: at high SNR levels, excessively large conditional sampling strengths cause reconstruction errors to increase. This finding suggests that although conditional sampling strength controls the weight of observational guidance in the reverse process, over-reliance on noisy gradients can amplify deviations in the sampling trajectory, ultimately leading to instability and distortion in the reconstructed responses. In contrast, DOM—owing to its simple structure and stable gradient estimation—exhibits less sensitivity to fluctuations in conditional sampling strength.

These results indicate that, in engineering scenarios with significant measurement noise, the conditional sampling strength should be carefully tuned to avoid assigning excessive weight to noisy observations. Moderately reducing the influence of observational constraints on the generative process can enhance both stability and accuracy of the reconstruction. The DPS framework demonstrates strong adaptability under varying SNR conditions: by adjusting the conditional sampling strength as a key hyperparameter, DPS dynamically adapts the reverse sampling strategy according to the reliability of the observations. This property highlights its robustness to noise and broad applicability in engineering practice.

## 4.4. Impact of DDIM-Accelerated Sampling

To further enhance the practical applicability of the DPS framework, it is necessary to introduce accelerated sampling strategies. Specifically, this study employs the DDIM as an acceleration scheme to reduce the number of sampling iterations, thereby decreasing inference latency and improving overall response speed.

For simplicity, a sparse sampling mechanism with fixed skip intervals is adopted. For example, in a model trained with $T = 500$ steps, the original sampling trajectory is $t \in [T, T-1, \ldots, 1, 0]$. By applying a DDIM skip interval of 10, the trajectory is sparsified to $\tau \in [T, T-10, \ldots, 10, 0]$, significantly improving sampling efficiency. However, although the DDIM strategy greatly reduces computational cost, its skip-step inference mechanism sacrifices part of the fine-grained information in the diffusion process, which may affect the final reconstruction accuracy of structural responses.

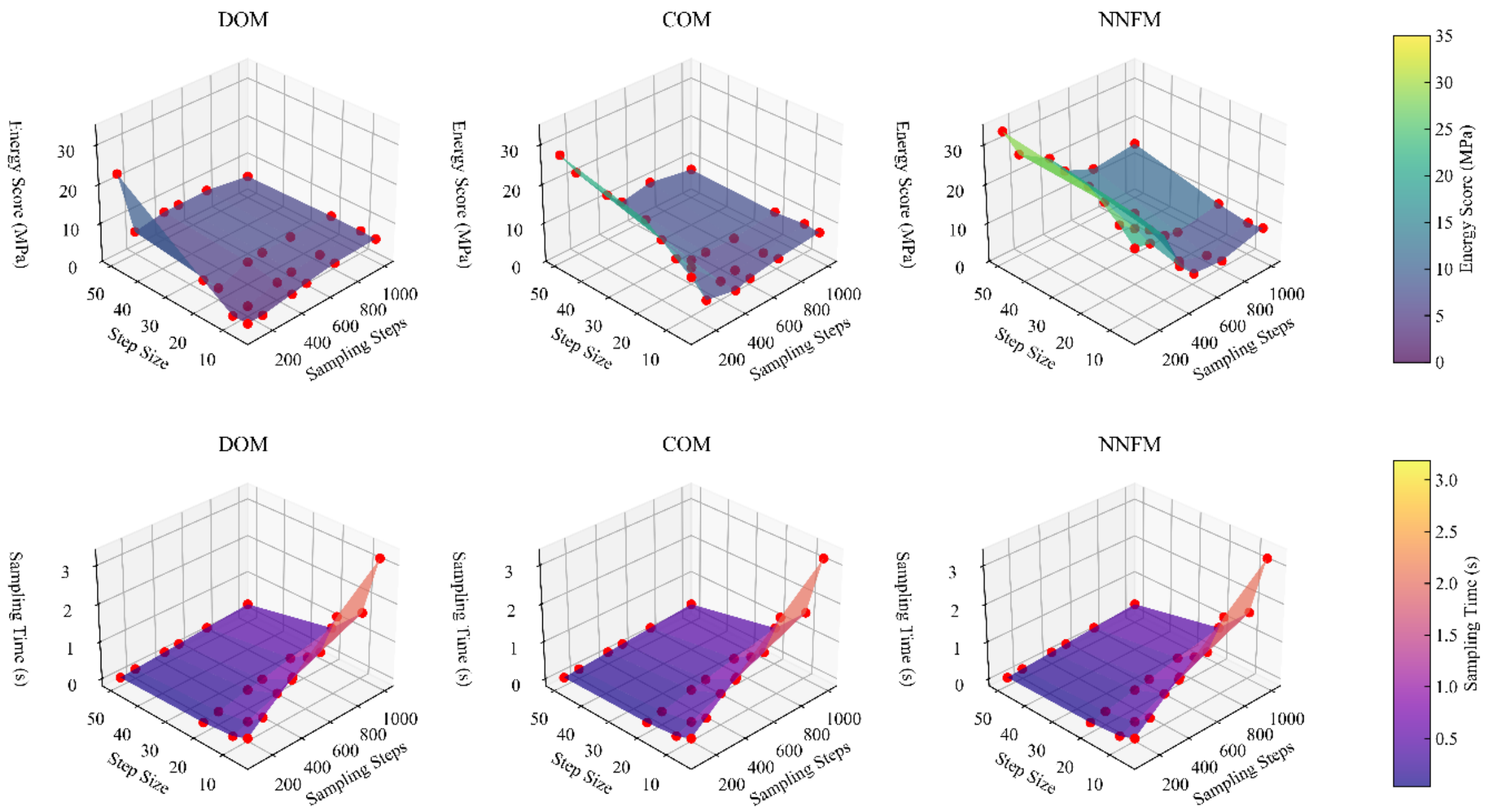

**Fig 14**. Reconstruction accuracy and inference time of different forward models under DDIM skip intervals and sampling steps

To evaluate the effect of different sampling steps and skip intervals on reconstruction performance, we conducted parallel sampling of 100 realizations under fixed observation conditions to simulate an ensemble prediction scenario. The average reconstruction accuracy and inference time were then calculated across multiple test samples, providing a joint evaluation of accuracy and efficiency. **Fig 14** presents the reconstruction accuracy (first row) and computational time (second row) of different forward models under various skip intervals and sampling step settings.

With respect to reconstruction accuracy, all three models exhibit a consistent trend: when the number of sampling steps is small, reconstruction accuracy is highly sensitive to the skip interval, with noticeable error fluctuations. As the number of steps increases, the influence of the skip interval diminishes, and the models demonstrate greater robustness to sparse sampling. This indicates that larger sampling steps improve stability in the generative process.

Regarding inference time, the three models show nearly identical computational costs under the same sampling settings, with overall time remaining low. Even under the densest configuration (1000 steps with a skip interval of 5), the reconstruction time remains below two seconds, underscoring the computational efficiency of diffusion models. Under balanced settings (e.g., 500 steps with a skip interval of 10), all three models complete the reconstruction of 100 parallel samples in under one second, fully satisfying the requirements of real-time applications such as structural health monitoring.

In summary, the reconstruction algorithms within the DPS framework achieve not only stable accuracy but also high computational efficiency. By properly configuring the sampling steps and skip intervals, it is possible to substantially reduce inference time while maintaining reconstruction accuracy, thereby meeting the dual demands of response speed and computational resource efficiency in engineering deployment.

## 5. Conclusion

This paper presents a novel data fusion framework utilizing diffusion models to address the challenge of reconstructing full-range structural responses from sparse and heterogeneous sensor measurements. The main conclusions are as follows:

(1) The full-range response reconstruction task fusing multi-sensors is formulated from the perspective of diffusion models. By employing DPS, the framework realizes conditional generation guided by conditional scores, which naturally supports the modeling requirement of recovering full-range responses from local measurements in SHM.

(2) Three forward models are developed under the DPS framework: DOM, COM, and NNFM. These models are tailored for different sensing scenarios, with DOM suited to consistent variable types and COM/NNFM handling heterogeneous sensor–response mappings. All three achieve good reconstruction accuracy on a systematic evaluation on a SPSW, with WMAPE of 1.62% (DOM), 3.27% (COM), and 3.49% (NNFM).

(3) A comprehensive evaluation under varying hyperparameters and sensor configurations shows that the framework can adapt to different application scenarios through adjustment of the conditional sampling strength. Moreover, owing to the flexibility of modifying forward models, the framework enables rapid assessment of alternative sensor placement strategies, offering potential for optimizing sensor deployment.

In summary, the proposed framework opens new possibilities for probabilistic modeling and sensor data fusion in structural health monitoring. Its capability for accurate full-range response reconstruction from sparse and noisy data paves the way for real-time structural estimation and enhanced infrastructure safety.

## Acknowledgements

The authors gratefully acknowledge the financial support provided by the National Natural Science Foundation of China (Grant No. 52408188, 52293433).

## CRediT authorship contribution statement

**Wingho Feng:** Data curation, Formal analysis, Methodology, Validation, Writing – original draft, Writing – review & editing, Visualization. **Quanwang Li:** Supervision, Project administration, Resources. **Chen Wang:** Conceptualization, Methodology, Writing – review & editing. **Jian-sheng Fan:** Project administration, Resources.